\documentclass[a4paper,fleqn,usenatbib]{mnras}
\usepackage{amsmath}
\usepackage{mathptmx}
\usepackage{txfonts}
\usepackage[T1]{fontenc}
\usepackage{ae,aecompl}
\usepackage{graphicx}	
\usepackage{amssymb}	
\usepackage[normalem]{ulem}
\newcommand{\be}{\begin{equation}}
\newcommand{\e}{\end{equation}}
\newcommand{\bear}{\begin{eqnarray}}
\newcommand{\ear}{\end{eqnarray}}

\def\aj{AJ}
\def\apj{ApJ}
\def\apjs{ApJS}
\def\jcap{JCAP}
\def\mnras{MNRAS}
\def\aap{A\&A}
\def\prd{Physical Review D}

\def\apjs{ApJS}
\def\apjl{ApJ Letters}

\title{}
\author{}
\title[Configuration entropy in the Universe] {Configuration entropy in the $\Lambda$CDM and the dynamical dark energy models: Can we distinguish one from the other?}
    
\author[Das, B. and Pandey, B.] {Biswajit
  Das\thanks{E-mail:bishoophy@gmail.com} and {Biswajit
    Pandey\thanks{E-mail: biswap@visva-bharati.ac.in}} \\ Department of
    Physics, Visva-Bharati University, Santiniketan, Birbhum, 731235,
    India\\ }
   
 \date{\today}

 \pubyear{2018}
  
\begin{document}
\label{firstpage}
\pagerange{\pageref{firstpage}--\pageref{lastpage}} 

\maketitle

\begin{abstract}
The evolution of the configuration entropy of the mass distribution in
the Universe is known to be governed by the growth rate of density
perturbations and the expansion rate of the Universe. We consider the
$\Lambda$CDM model and a set of dynamical dark energy models with
different parametrization of the equation of state and explore the
evolution of the configuration entropy in these models. We find that
the nature of evolution of the configuration entropy depends on the
adopted parametrization which may allow us to discern them from each
other. The configuration entropy initially decreases with time but
nearly plateaus out at present in the $\Lambda$CDM model. The models
where dark energy becomes less dominant at late times exhibit a larger
decrease in the configuration entropy compared to the $\Lambda$CDM
model after redshift $z \sim 1$. We find that the configuration
entropy remains nearly unchanged in the models where dark energy
becomes more dominant at late times.  Our results suggest that the
method presented here may be also used to constrain the initial value
of the configuration entropy of the Universe.

\end{abstract}

 \section {Introduction}
 Understanding the observed accelerated expansion of the Universe
 remains one of most interesting and challenging problems in
 cosmology. The fact that we live in an expanding Universe is quite
 well known since Edwin Hubble's milestone discovery of the Hubble's
 law \citep{hubble}. The observational evidence of accelerating
 expansion of the Universe was confirmed much later by two independent
 groups \citep{riess, perlmutter1} in 1998. This proved to be a
 landmark discovery in cosmology which revolutionized our current
 understanding of the Universe. The accelerated expansion of the
 Universe demands an explanation given the attractive nature of
 gravity and presence of matter in the Universe. The most common
 hypothetical explanation at present is that there exists something
 called dark energy which has negative pressure and whose energy
 density remains constant despite the expansion of the Universe. This
 hypothetical entity is very often identified with the cosmological
 constant of Einstein's field equation but the question of the
 physical origin of this component is far from being
 settled. Different physically motivated ideas such as the
 backreaction mechanism \citep{buchert2k}, effect of a large local
 void \citep{tomita01, hunt}, entropic force \citep{easson},
 information storage in the spacetime \citep{paddy, paddyhamsa} have
 been proposed to explain the accelerated expansion of the Universe. A
 number of various other alternative models such as quintessence
 \citep{ratra, caldwell}, k-essence \citep{armendariz},
 extra-dimensional models \citep{milton}, and modified gravity
 \citep{tsujikawa} have been also proposed to understand the physical
 origin and the nature of dark energy. A thorough discussion on these
 models and the possible ways to constrain them can be found in
 \citet{copeland} and \citet{de2010}.

The time-independent cosmological constant is regarded as the simplest
dark energy candidate and the corresponding model is the $\Lambda$CDM
model which has been proved to be the most successful model in
explaining a wide range of cosmological observations. Despite its
great success in explaining many cosmological observations, the
$\Lambda$CDM model is still plagued by the cosmological constant
problem and coincidence problem in the context of dark energy. In an
attempt to construct more natural models, a particular class of models
known as the dynamical dark energy models with time-dependent equation
of state have been proposed \citep{caldwell}. If the equation of state
of dark energy varies with time, it is argued that the signature of
that variation would be found in the expansion history and the growth
of the large scale structures in the Universe.
 
A number of other studies \citep{radicella, pavon1, mimoso, pavon2,
  ferreira} suggest that the accelerated expansion of the Universe may
be related to the second law of thermodynamics. It has been shown that
the entropy of the Universe tends to some maximum value in the
$\Lambda$CDM model \citep{pavon2} whereas different other models such
as the modified gravity theories, nonsingular bouncing Universes and
the Universes dominated by matter or phantom fields do not tend to a
state of maximum entropy \citep{radicella, mimoso, ferreira}. 

The entropy of the relativistic particles remains unchanged during the
expansion of the Universe. The growth of the Stellar Black Holes (SBH)
and the Supermassive Black Holes (SMBH) increases the entropy of the
Universe
\citep{bekenstein,hawking,penrose,frampton,egan}. \citet{gibbons} show
that besides other sources, an entropy is associated with the
Cosmic Event Horizon (CEH) which is the most dominant source of
entropy in the Universe \citep{egan}.

In a recent work, \citet{pandey1} proposed that the transition of the
Universe from a nearly smooth to a highly clumpy state due to the
growth of structures by the gravitational instability is associated
with a large decrease in the configuration entropy of the Universe.
It has been argued that the present acceleration of the Universe may
arise to counterbalance this dissipation of the configuration
entropy. It is well known that a static Universe with gravity and
matter is unstable. The growth of structure at small scales would
force the configuration entropy to decrease. If we treat the Universe
as a thermodynamic system, then the second law of thermodynamics
should hold for the Universe as a whole. Since none of the known
entropy generation processes are efficient enough to counter the loss
of configuration entropy, the Universe must expand to prevent the
further growth of structures. The configuration entropy continues to
dissipate in a matter dominated Universe. Interestingly, this
dissipation is damped out in a $\Lambda$ dominated Universe where the
entropy rate becomes zero and the configuration entropy tends to some
maximum value.

In the present work, we explore the behaviour of the configuration
entropy in the $\Lambda$CDM model and a set of dynamical dark energy
models to test if they can be distinguished from each other based on
the configuration entropy. The Universe in the radiation dominated era
was very smooth and hence possessed very high configuration entropy on
all scales. The relative inhomogeneities observed in the CMBR is of the
order of $10^{-5}$ at the time of recombination. These inhomogeneities
were amplified further by the gravitational instability leading to the
growth of large scale structures in the Universe. As the structures
form, the universe becomes more clumpy or inhomogeneous leading to a
larger dissipation of the configuration entropy. The configuration
entropy rate is known to depend on the growth rate of structures and
the expansion rate of the Universe both of which in turn depend on the
cosmological model under consideration. We study the consequences of
the dynamical nature of the equation of state on the growth rate and
the expansion rate and subsequently on the configuration entropy of
the Universe. Configuration entropy in the present Universe can be
easily measured from the large redshift surveys like SDSS
\citep{york}. In future, combining measurements from future galaxy
surveys like DESI and Euclid and observations from the different
future $21$ cm experiments would enable us to measure the
configuration entropy of the Universe at different redshifts. This
would then allow us to constrain the various cosmological parameters
by measuring the configuration entropy of the Universe and its
evolution.
  
  \section {Theory}
  \subsection {Evolution of configuration entropy}
  We consider a significantly large comoving volume $V$ of the
  Universe over which the Universe can be treated as homogeneous and
  isotropic. This large volume is further subdivided into smaller
  sub-volumes $dV$. Let $\rho (\vec{x},t)$ be the density measured
  inside the different subvolumes. Here $\vec{x}$ denote the comoving
  coordinate assigned to each sub-volumes.

\citet{pandey1} defines the configuration entropy following the
definition of information entropy \citep{shannon48} as
  \begin {eqnarray}
   S_c(t) = - \int \rho \log \rho \, dV.
   \label{eq:one}
  \end {eqnarray}

  The continuity equation for a fluid in an expanding universe in
  comoving coordinate can be expressed as,
  \begin {eqnarray}
   \frac {\partial \rho}{\partial t} + 3 \frac {\dot a}{a}\rho + \frac
         {1}{a}\nabla \cdot (\rho \vec {v}) = 0.
   \label{eq:two}
  \end {eqnarray}
  where $a$ is the scale factor and $\vec {v}$ is the peculiar
  velocity of the fluid element contained in a subvolume $dV$.

  Multiplying \autoref{eq:two} by $(1 + \log \rho)$ and integrating
  over the entire volume $V$, we get \citep{pandey1},
  \begin {eqnarray}
   \frac {dS_c(t)}{dt} + 3\frac{\dot a}{a}S_c(t) - \frac{1}{a}\int \rho (3 \dot a + \nabla \cdot \vec {v})\, dV = 0.
   \label{eq:three}
  \end {eqnarray}
  Redefining $F(t)=\frac{1}{a}\int \rho (3 \dot a + \nabla \cdot \vec
  {v})\, dV$ and changing the variable from $t$ to $a$, the
  \autoref{eq:three} becomes
  \begin {eqnarray}
   \frac {dS_c(a)}{da}\dot a + 3\frac {\dot a}{a}S_c(a) - F(a) = 0.
   \label{eq:four}
  \end {eqnarray}
where $F(a)$ can be expressed as,
\begin{eqnarray}
   F(a) = 3MH(a) + \frac {1}{a} \int \rho (\vec {x}, a)\nabla \cdot \vec {v}\, dV.
   \label {eq:five}
  \end{eqnarray}
 
  Here $M = \int \rho (\vec {x}, a)\, dV = \int \bar \rho (1 + \delta
  (\vec {x}, a))\, dV$ is the total mass contained inside the comoving
  volume $V$. $\delta (\vec {x}, a) = \frac {\rho (\vec {x}, a) - \bar
    \rho}{\bar \rho}$ is the density contrast and $\bar \rho$ is the
  mean density of matter within the comoving volume under
  consideration.
  
  In linear perturbation theory, the density perturbations evolve as,
  $\delta (\vec x, a) = D(a)\delta (\vec x)$ where $D(a)$ is the
  growing mode of density perturbations. The divergence of the
  peculiar velocity $\vec{v}$ can be expressed as,

  \begin {eqnarray}
   \nabla\cdot \vec v = - a \frac {\partial \delta(\vec {x}, t)}{\partial t} = - a\frac {\partial \delta(\vec {x}, a)}{\partial a}\dot a = - a \dot a \frac {dD(a)}{da}\delta(\vec x) .
   \label {eq:six}
   \end {eqnarray}
  
Writing $\rho(\vec {x}, a)$ as $\bar\rho(1+\delta(\vec {x}, a))$ and
inserting the expression for $\nabla\cdot \vec v$ in \autoref{eq:four}
we get,

\begin {eqnarray}
  \frac {dS_c(a)}{da}\dot a + 3 \frac {\dot a}{a}(S_c(a) - M)\nonumber \\ 
  + \frac{1}{a}\bar \rho a\dot a\frac{dD(a)}{da}\Bigg[\int \delta(\vec x)\, dV + D(a)\int \delta^2(\vec x)\, dV\Bigg] = 0.
   \label{eq:seven}
  \end {eqnarray}
  The first term in the square bracket is zero by definition. So we
  finally have an ordinary first-order differential equation,
  \begin {eqnarray}
  \frac {dS_c(a)}{da} + \frac {3}{a}(S_c(a) - M) + \bar\rho f\frac {D^2(a)}{a}\int \delta^2(\vec x)\, dV = 0.
  \label {eq:eight}
\end {eqnarray}
where, $f = \frac {dlnD}{dlna} = \frac {a}{D} \frac {dD}{da}$ is the
dimensionless linear growth rate.

We have to solve \autoref {eq:eight} to determine the evolution of the
configuration entropy in different cosmological models. We set the
time independent quantities in the third term of \autoref {eq:eight}
to be $1$ for the present analysis. We then calculate $D(a)$ and $f$
for different cosmological models considered in this work and
numerically solve \autoref {eq:eight} using the 4th order Runge-Kutta
method.

It is interesting to note that the evolution of the configuration
entropy is governed by the initial condition at the beginning when the
contribution of the third term is negligible as compared to the second
term in \autoref{eq:eight}. The cosmology dependence of $S_c(a)$
purely arises from the third term which involves the growth rate and
its derivative. An analytical solution of \autoref{eq:eight} ignoring
the third term in it is given by, 
\begin {eqnarray}
  \frac {S_c(a)}{S_c(a_i)} = \frac{M}{S_c(a_i)}+\Bigg(1-\frac{M}{S_c(a_i)}\Bigg)\Bigg(\frac{a_i}{a}\Bigg)^3.
  \label {eq:nine}
\end {eqnarray}
 where, $a_i$ is the initial scale factor which we have
  chosen to be $10^{-3}$ throughout the analysis.

\subsection{Growing mode and the dimensionless linear growth rate of density perturbation}
 
Any primordial density perturbations present in the early Universe
will be amplified by the gravitational instability. Initially the
density of matter was slightly higher than the average in some regions
whereas it was a little lower in some other regions. The overdense
regions would turn into even denser regions and the underdense regions
would be more underdense with time. The growth of the density
perturbations can be studied reliably with the linear perturbation
theory when the density perturbations $\delta<<1$. If we consider only
the perturbations to the matter then the growth equation becomes,
  \begin {eqnarray}
   \ddot \delta(t) + 2H \dot \delta(t) - (3/2)H^2\Omega_m \delta(t) = 0.
    \label{eq:thirteen}
  \end {eqnarray}
  Changing the variable of differentiation from $t$ to $a$ the \autoref
  {eq:thirteen} becomes \citep{linder4}
  \begin {eqnarray}
   \delta^{\prime\prime}(a) + \Big(\frac {2-q}{a}\Big)\delta^{\prime}(a) - \frac {3\Omega_m}{2a^{2}}\delta(a) = 0,
   \label{eq:fourteen}
  \end {eqnarray}
  where a prime over $\delta$ means derivative with respect to $a$,
  $\Omega_m$ is the density parameter and $q=\frac {-a \ddot a}{\dot
    a^{2}}$ is the deceleration parameter.
  
If we assume that the universe has only two components such as matter
(baryonic matter and dark matter) and dark energy whose equation of
state is parametrized with some function $\omega(a)$ then we can
write,
  \begin {eqnarray}
  \frac {H^2(a)}{H^2_0} = E^2(a)=\Omega_m a^{-3} + (1 - \Omega_m)e^
        {3\int_a^1 [1+\omega(a^{\prime})]\, d \ln a^{\prime}}.
   \label{eq:sixteen}
  \end {eqnarray}
  Here $H_0$ is the present value of the Hubble parameter. The time
  varying nature of the dark energy is encoded in the parametrization
  of $\omega(a)$.  This may appear as ad-hoc and phenomenological but
  it is more generic approach \citep{linder4}. We rewrite \autoref{eq:fourteen} as
  \citep{linder4}
  \begin{eqnarray}
   D^{\prime\prime} + \frac
   {3}{2}\Bigg[1-\frac{\omega(a)}{1+X(a)}\Bigg]\frac{D^{\prime}}{a} -
   \frac {3}{2}\frac{X(a)}{1+X(a)}\frac{D}{a^2} = 0.
   \label{eq:seventeen}
  \end{eqnarray}
  where $D = \frac{\delta(a)}{\delta(a_i)}$ and $\delta(a_i)$ is
  density contrast at some initial scale factor $a_i$ and
  \begin{eqnarray}
   X(a) = \frac{\Omega_m}{1-\Omega_m}e^{-3\int_a^1 \omega(a^{\prime})\, d \ln a^{\prime}}.
   \label{eq:eighteen}
  \end{eqnarray}
  \autoref {eq:seventeen} explicitly shows that the growth rate depends
  on the functional form of $\omega(a)$ and will give different result
  when a different $\omega(a)$ is chosen. We numerically solve the
  \autoref{eq:seventeen} using the 4th order Runge-Kutta method. We
  normalize the solution $D(a)$ so that $D(a_0) =1$ in the
  $\Lambda$CDM model where $a_{0}$ is the present scale factor.

  We can write the matter density history as,
  \begin{eqnarray}
   \Omega_m(a) = \frac{\Omega_ma^{-3}}{E^2(a)}.
   \label{eq:nineteen}
  \end{eqnarray}
  and the dimensionless linear growth rate as
  \begin{eqnarray}
   f = [\Omega_m(a)]^{\gamma},
   \label{eq:twenty}
  \end{eqnarray}
  where $\gamma$ is the growth index which can be obtained by using
  the fitting formula provided by \citet{linder2} as,
  \begin{eqnarray}
   \gamma = 0.55 + 0.05[1+\omega(z=1)].
   \label{eq:twentyone}
  \end{eqnarray}
  
We estimate the dimensionless linear growth rate $f$ by combining
\autoref{eq:nineteen}, \autoref{eq:twenty} and \autoref{eq:twentyone}.

\begin{figure*}
   \resizebox{7 cm}{!}{\rotatebox{0}{\includegraphics{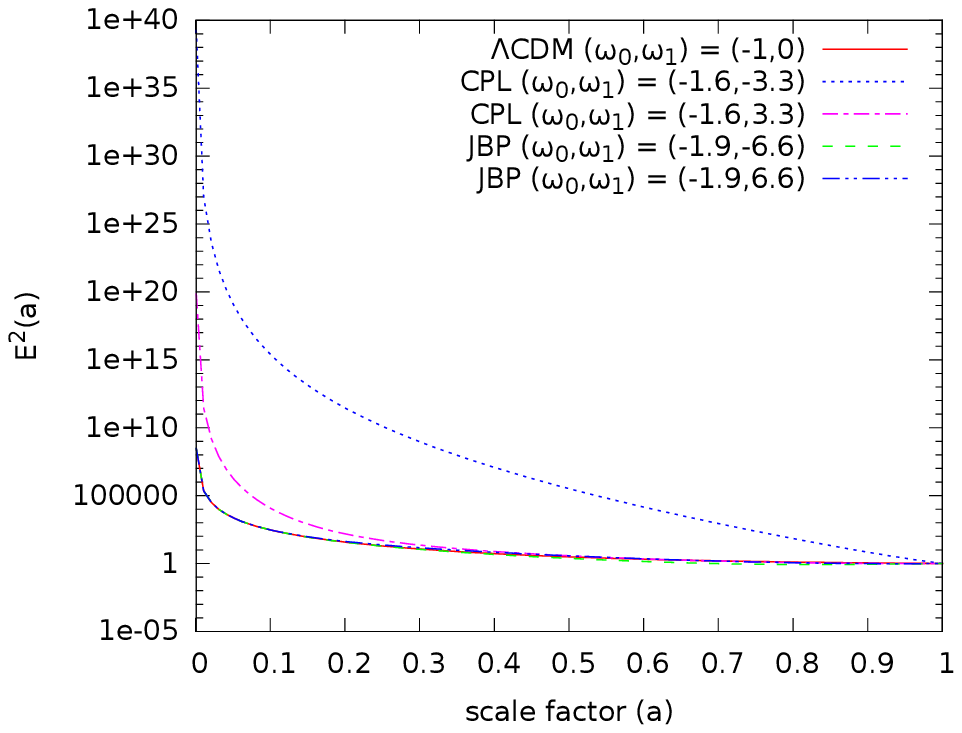}}}
   \hspace{0.5 cm}
   \resizebox{7 cm}{!}{\rotatebox{0}{\includegraphics{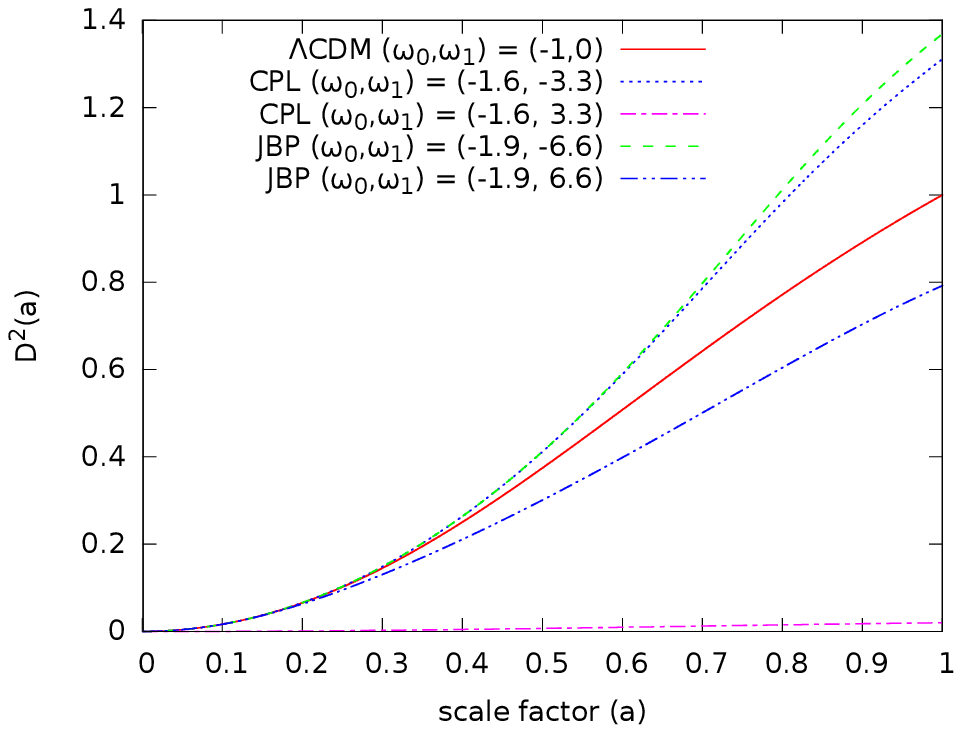}}}\\
   \vspace{-0.1 cm} 
   \resizebox{7 cm}{!}{\rotatebox{0}{\includegraphics{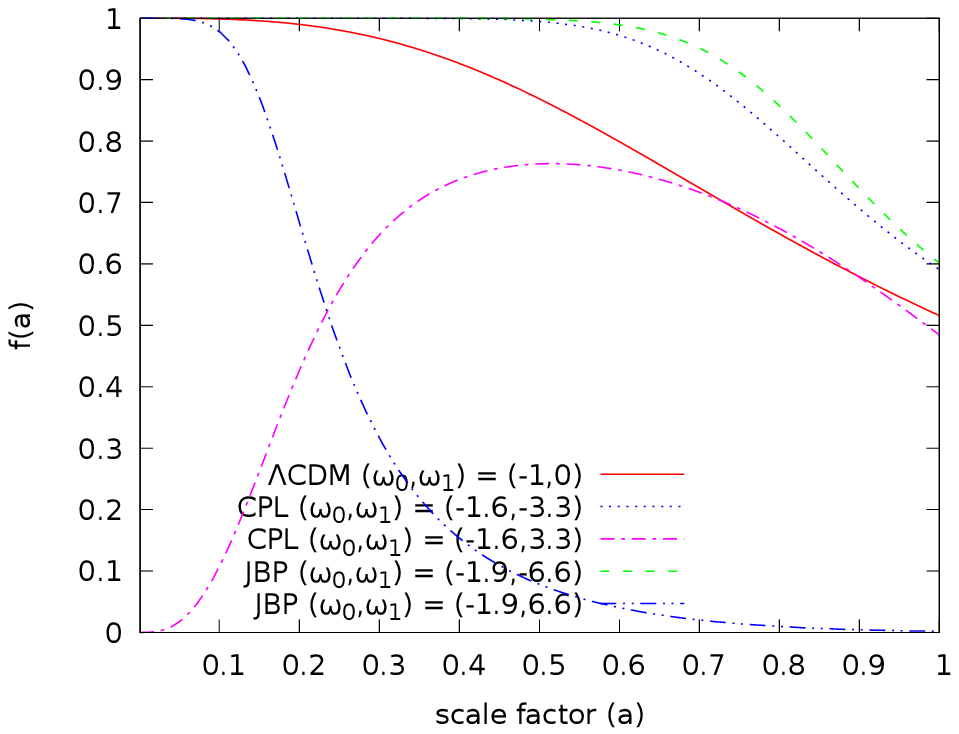}}}
   \hspace{0.5 cm}
   \resizebox{7 cm}{!}{\rotatebox{0}{\includegraphics{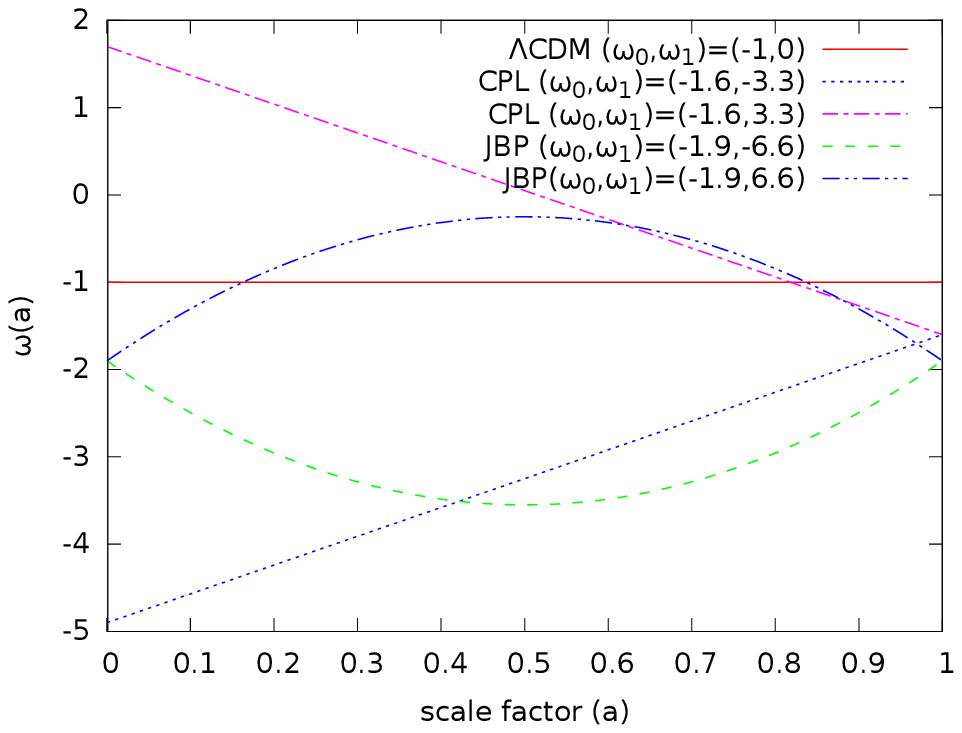}}}\\
   \caption{The top left panel shows the evolution of $E^2(a)$ for CPL
     and JBP parametrization of dark energy along with that for the
     $\Lambda$CDM model. The top right panel shows the evolution of
     $D^2(a)$ for the same models. Both $E^2(a)$ and $D^2(a)$ are
     normalized to 1 at present in the $\Lambda$CDM model. The
     respective values for the other models are normalized with
     respect to the $\Lambda$CDM model. The bottom left and right
     panels show the evolution of the dimensionless linear growth rate
     and the equation of state parameter in these models
     respectively.}
   \label{fig:e2d2fwall}
  \end{figure*}

 \begin{figure*}
   \resizebox{7 cm}{!}{\rotatebox{0}{\includegraphics{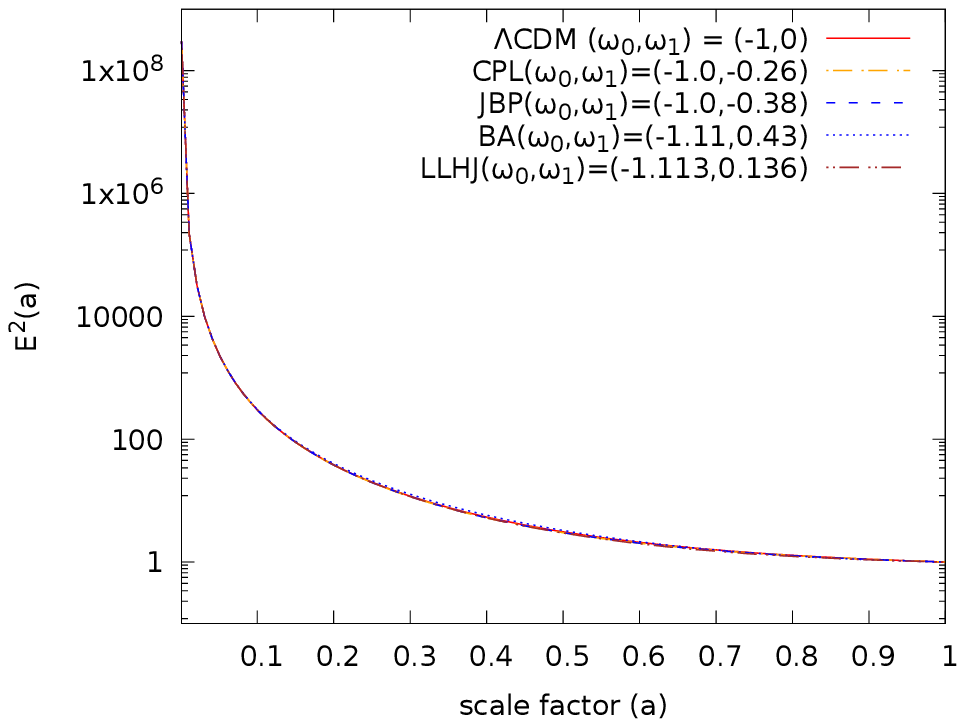}}}
   \hspace{0.5 cm}
   \resizebox{7 cm}{!}{\rotatebox{0}{\includegraphics{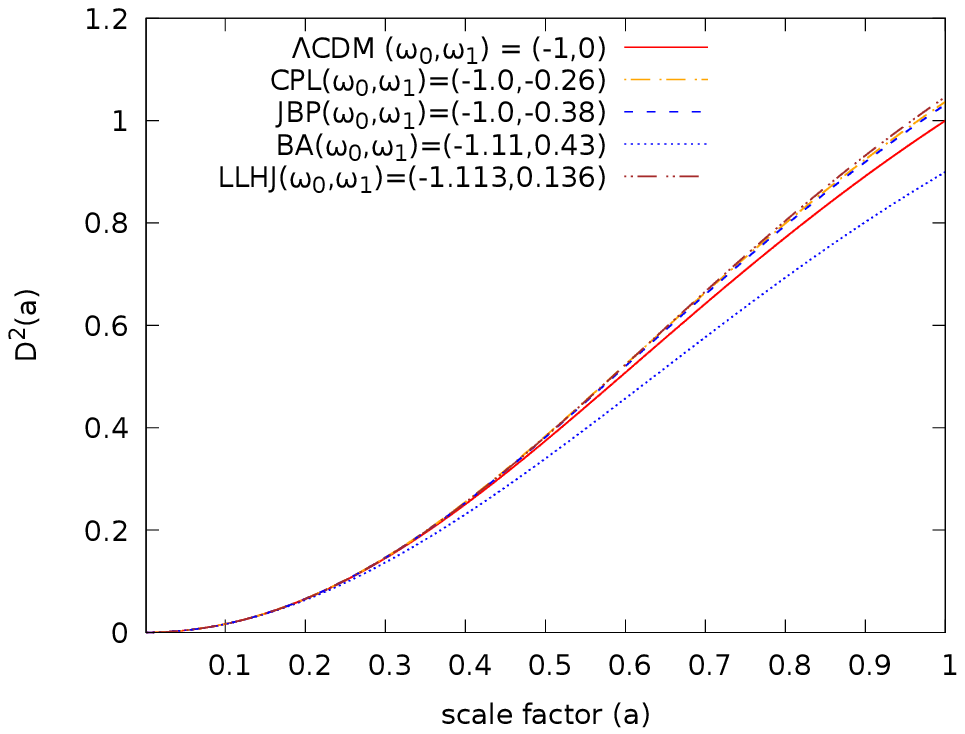}}}\\
   \vspace{-0.1 cm} 
   \resizebox{7 cm}{!}{\rotatebox{0}{\includegraphics{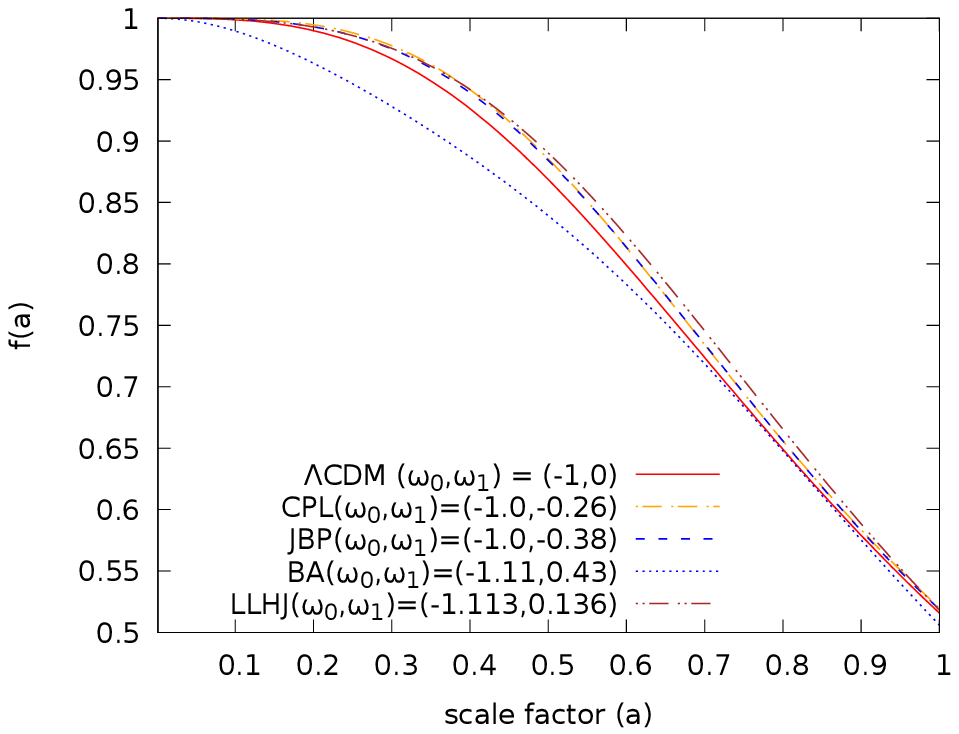}}}
   \hspace{0.5 cm}
   \resizebox{7 cm}{!}{\rotatebox{0}{\includegraphics{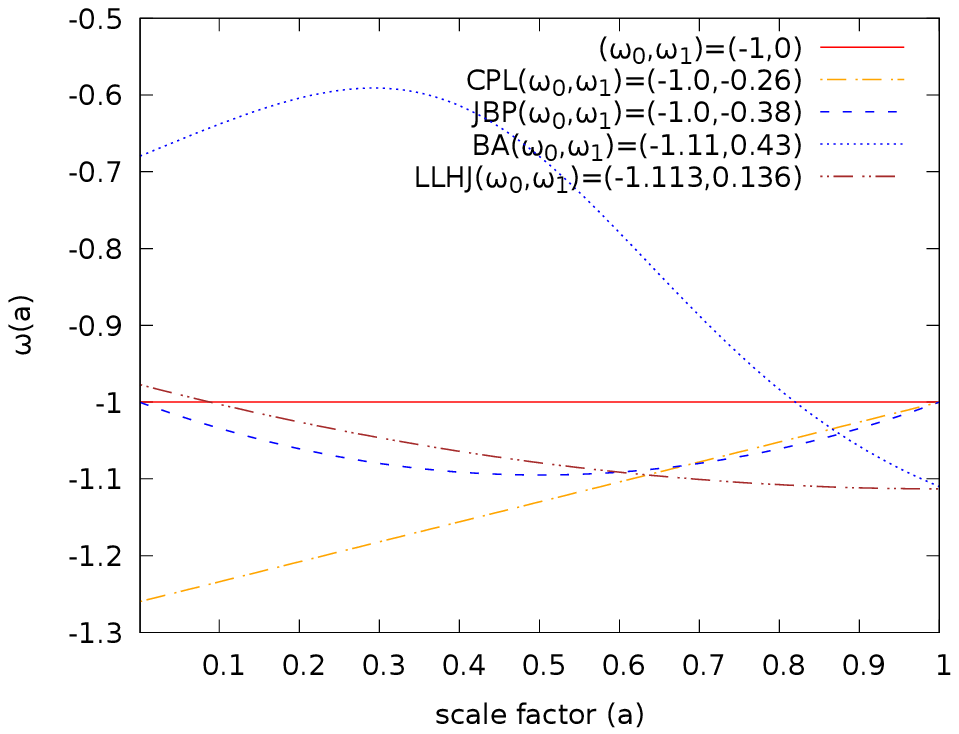}}}\\
   \caption{Same as \autoref{fig:e2d2fwall} but for BA and LLHJ
       parametrization together with CPL and JBP parametrization with
       different choice of parameters.}
   \label{fig:e2d2fwnew}
  \end{figure*}
  
 \subsection{Different parametrization of dynamical dark energy models}
  A number of two-parameter descriptions of $\omega$ have been
  explored in the literature. In the present work, we consider a
  number of different parametrization of dynamical dark energy. For
  each of these models we calculate the growing mode of density
  perturbation $D(a)$ and the dimensionless linear growth rate $f$
  using the equations outlined in the subsection 2.2. We then solve
  \autoref{eq:eight} to study the evolution of the configuration
  entropy in these models.

  \subsubsection{Chevallier-Polarski-Linder (CPL) Parametrization}
  In CPL parametrization \citep{chevallier,linder1}, the equation of
  state is parametrized as,
  \begin{eqnarray}
  \omega(a) = \omega_0 + \omega_1(1-a).
  \label{eq:twentythree}
  \end{eqnarray}
  The main advantage offered by this parametrization is its bounded
  behaviour at infinite redshift. The slope of the equation of state
  $\frac{d\omega(a)}{da} = -\omega_1$ is constant and the sign of the
  constant determines whether the energy density increases or
  decreases with scale factor.  

In the present work, we choose the values of the two constants
$\omega_0$ and $\omega_1$ from \citet{gong} and \citet{johri} who
claim that $\omega_0 = -1.6$, $\omega_1 = 3.3$ to be the best fit
values to SN Ia `gold set' data. We also consider the case for
$\omega_1 = -3.3$ following \citet{mamon} who suggest that the model
with negative $\omega_1$ is thermodynamically stable.
\subsubsection{Jassal-Bagla-Padmanabhan (JBP) Parametrization}
\citet{jbp} parametrized the equation of state as,
  \begin{eqnarray}
  \omega(a) = \omega_0 + \omega_1a(1-a) = \omega_0 + \omega_1a - \omega_1a^2.
  \label{eq:twentysix}
  \end{eqnarray}
  The slope of the equation of state $\frac{d\omega(a)}{da} =
  \omega_1(1-2a)$ is not a constant in this case. We use the best fit
  values to the SN Ia `gold set' data provided in \citet{gong} and
  \citet{johri} as $\omega_0 = -1.9$, $\omega_1 = 6.6$. We also
  consider the case for $\omega_1 = -6.6$ as before.

We also consider some models where $\omega(a)$ does not deviate much
from $-1$ throughout the entire range of evolution. We again consider
the CPL and JBP parametrizations albeit with some different choice of
parameters. We use ($\omega_0=-1.0,\omega_1=-0.26$) for the CPL and
($\omega_0=-1.0,\omega_1=-0.38$) for the JBP parametrization which are
constrained by \citet{tripathi} using SNIa+BAO+H(z) data. Two other
parametrizations \citep{barboza,liu} for which $\omega(a)$ remains
close to $-1$ for the entire redshift range are also considered.

\subsubsection{Barboza-Alcaniz (BA) Parametrization} The
  equation of state proposed in \citet{barboza} is given by,
\begin{eqnarray}
\omega(a) = \omega_0 + \omega_1\frac{(1-a)}{(2a^2-2a+1)}
\label{eq:twentyseven}
\end{eqnarray}

This parametrization also exhibit a bounded behaviour at infinite
redshift. We use the parameters $(\omega_0=-1.11, \omega_1=0.43)$ as
constrained by \citet{barboza} using SNIa, BAO, CMB and H(z) data.

\subsubsection{Liu-Li-Hao-Jin (LLHJ) Parametrization}
\citet{liu} propose a family of parametrizations. Family I is given by 
 \begin{eqnarray}
\omega(a) = \omega_0 + \omega_1 \left(1-a\right)^n
\label{eq:twentyeight}
\end{eqnarray}
Family II is given by  
\begin{eqnarray}
\omega(a) = \omega_0 + \omega_1 a^{n-1}(1-a)
\label{eq:twentynine}
\end{eqnarray}

Both family I and family II reduce to CPL parametrization for
$n=1$. For $n = 2$, family II reduces to JBP parametrization. We
consider the parametrization of family I with $n = 2$ in this
work. This parametrization starts with $\omega > -1$ but quickly
slides below $-1$. The values of the parameters $(\omega_0=-1.113,
\omega_1=0.136)$ are constrained by \cite{liu} using SNIa data.

The equation of state $\omega(a)$ comes into play while evaluating
$E^2(a)$ in \autoref{eq:sixteen} and $X(a)$ in \autoref{eq:eighteen}.
We show the evolution of $E^2(a)$, $D^2(a)$, the dimensionless growth
rate $f$ and the equation of state parameter $\omega(a)$ for different
cosmological models considered in the present work in the different
panels of \autoref{fig:e2d2fwall} and \autoref{fig:e2d2fwnew}.

 \section{Results and Conclusions}
 
  \begin{figure*}
   \resizebox{7 cm}{!}{\rotatebox{0}{\includegraphics{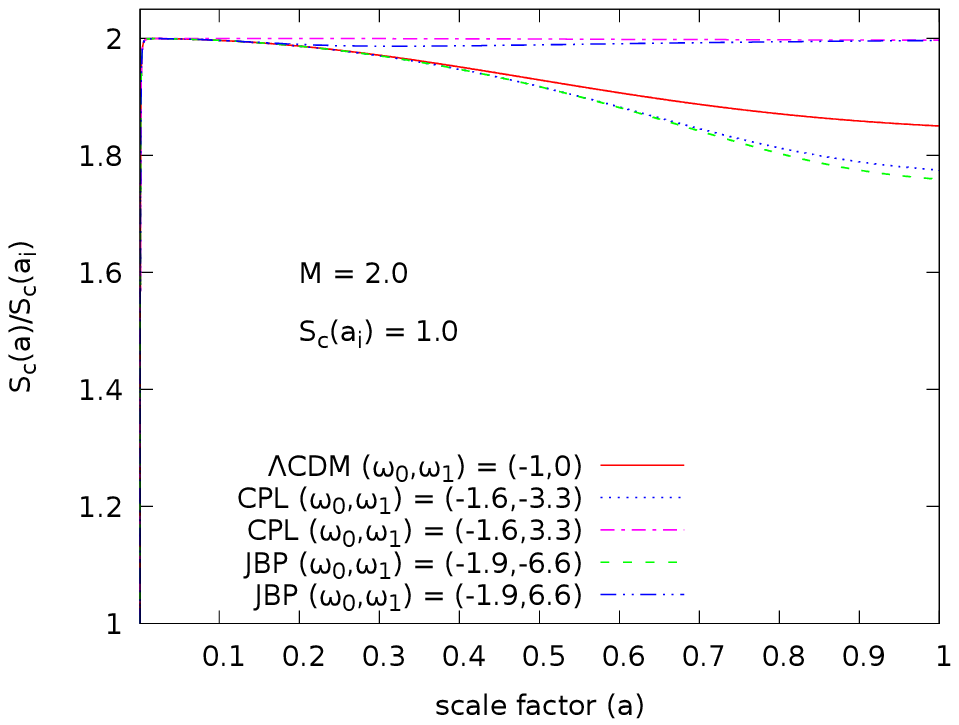}}}
   \hspace{0.5 cm}
   \resizebox{7 cm}{!}{\rotatebox{0}{\includegraphics{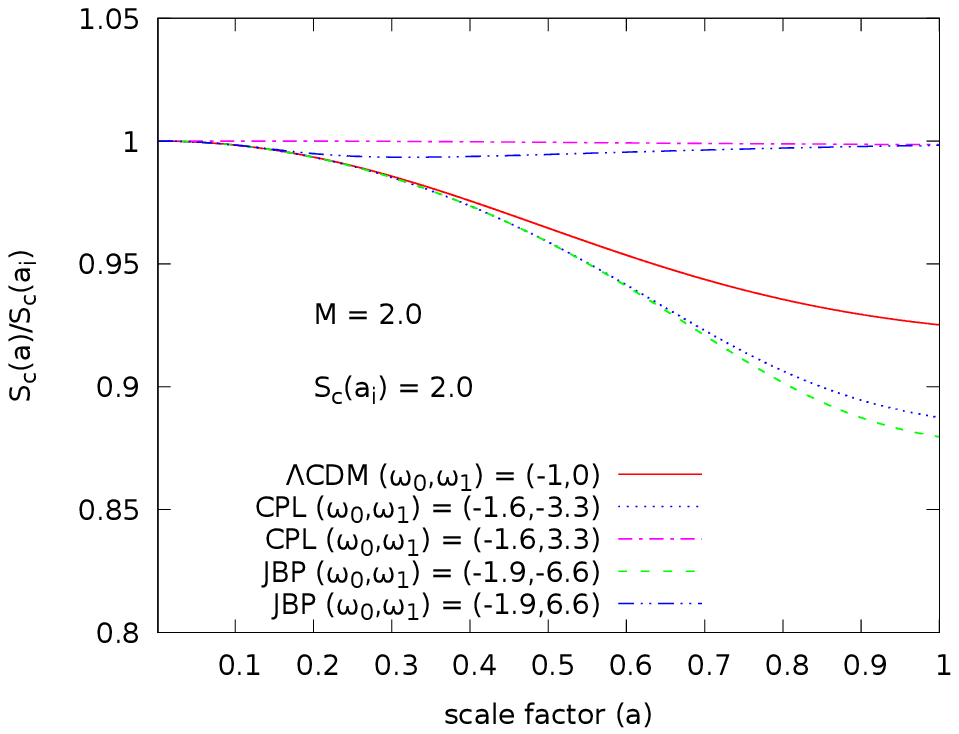}}}\\
   \vspace{-0.1 cm} 
   \resizebox{7 cm}{!}{\rotatebox{0}{\includegraphics{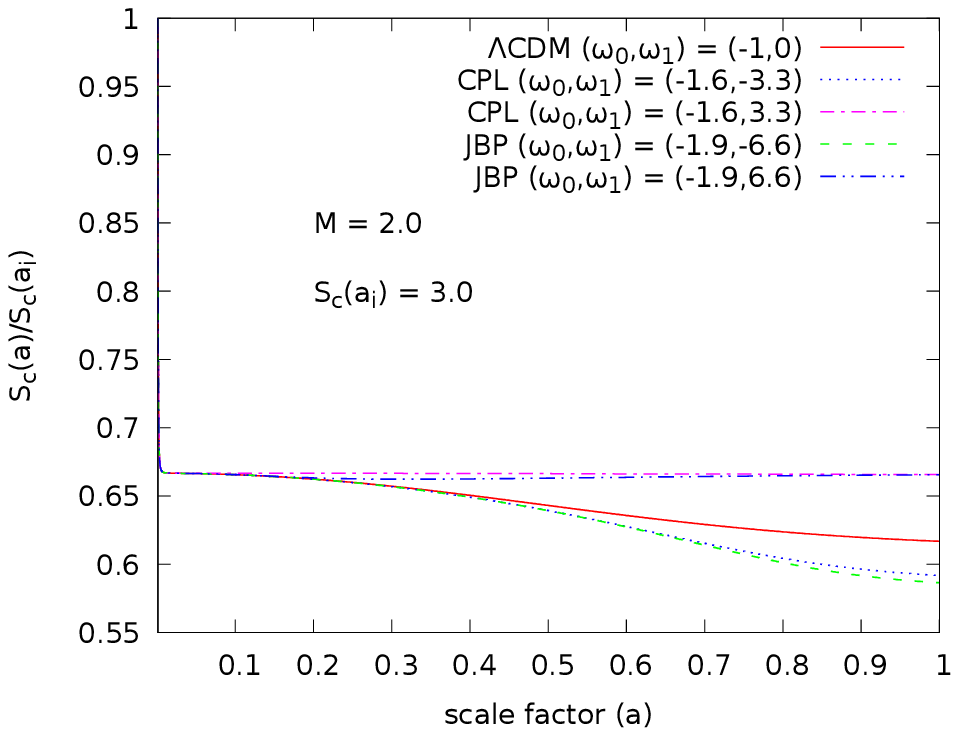}}}\\
   \vspace{-0.1 cm}
   \caption{Different panels show the evolution of the ratio of the
     configuration entropy $S_c(a)$ to its initial value $S_c(a_i)$
     for CPL and JBP parametrization of dark energy along with the
     $\Lambda$CDM model for different initial entropy $S_c(a_i)$ and
     mass $M$ enclosed within comoving volume V as indicated in each
     panel.}
   \label{fig:ent}
  \end{figure*}

 \begin{figure*}
   \resizebox{7 cm}{!}{\rotatebox{0}{\includegraphics{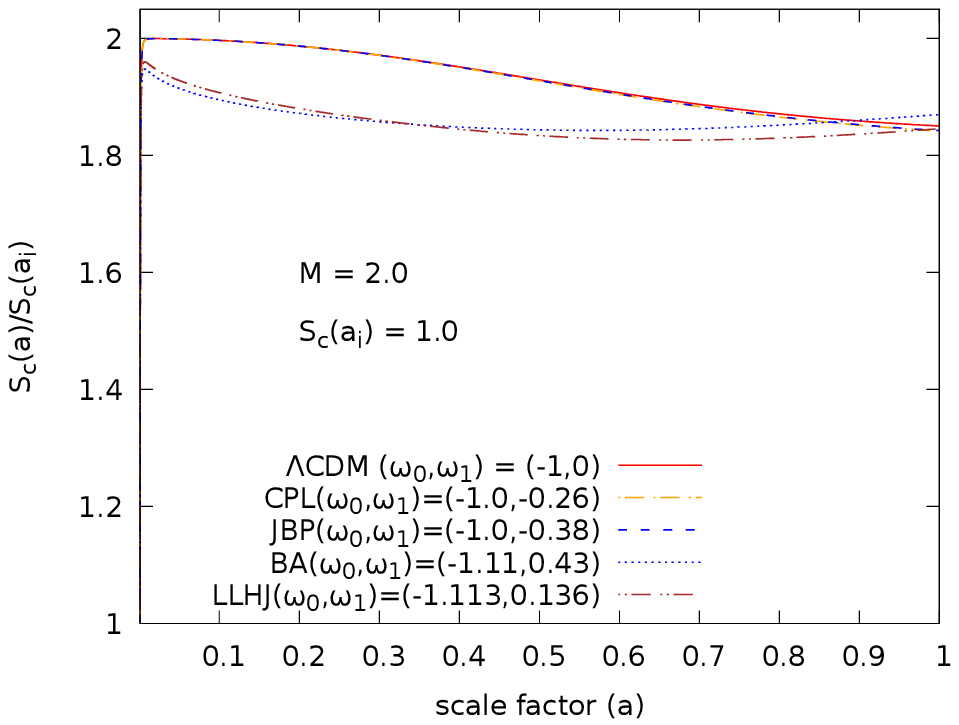}}}
   \hspace{0.5 cm}
   \resizebox{7 cm}{!}{\rotatebox{0}{\includegraphics{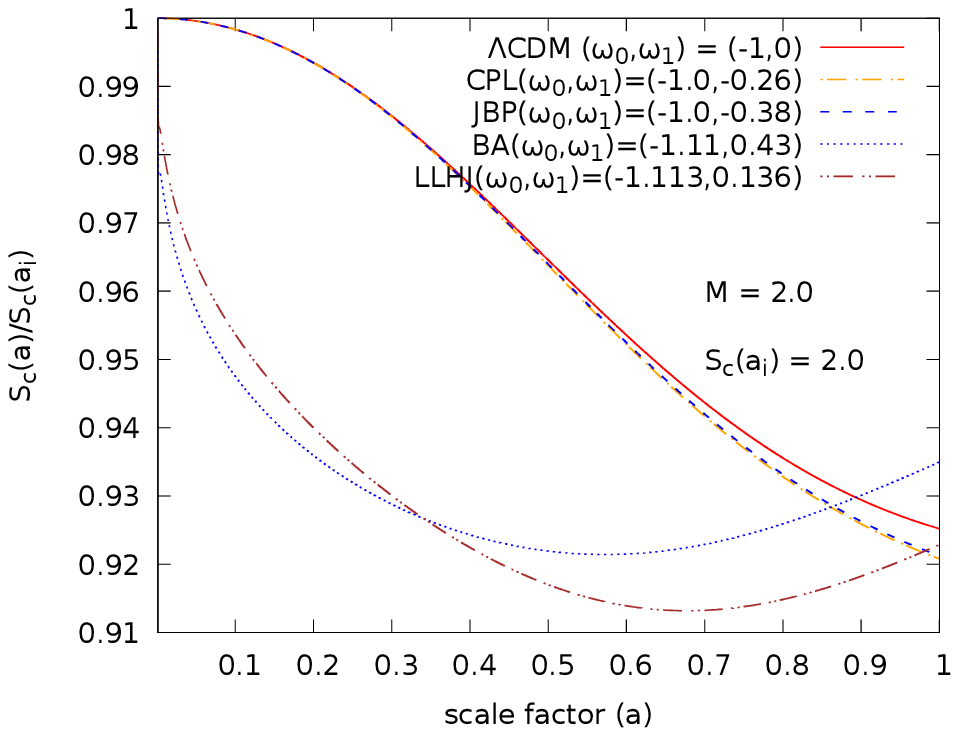}}}\\
   \vspace{-0.1 cm} 
   \resizebox{7 cm}{!}{\rotatebox{0}{\includegraphics{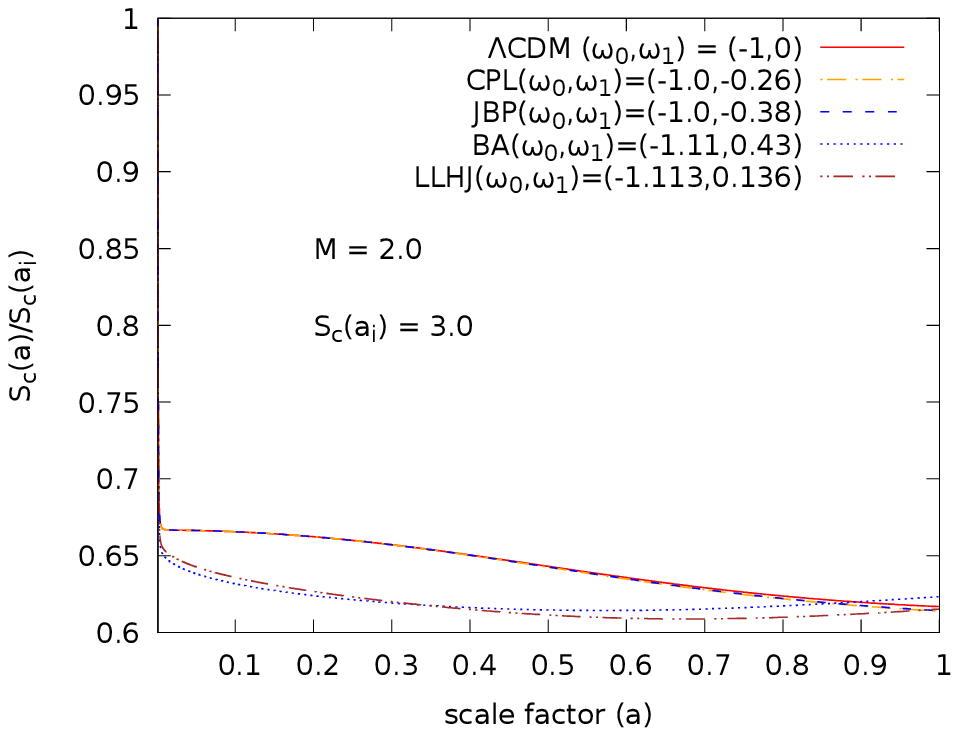}}}\\
   \vspace{-0.1 cm}
   \caption{ Same as \autoref{fig:ent} but for BA and LLHJ
     parametrization along with CPL and JBP parametrization with
     different choice of parameters. }
   \label{fig:entnew}
  \end{figure*}
  
 \begin{figure*}
 \resizebox{7 cm}{!}{\rotatebox{0}{\includegraphics{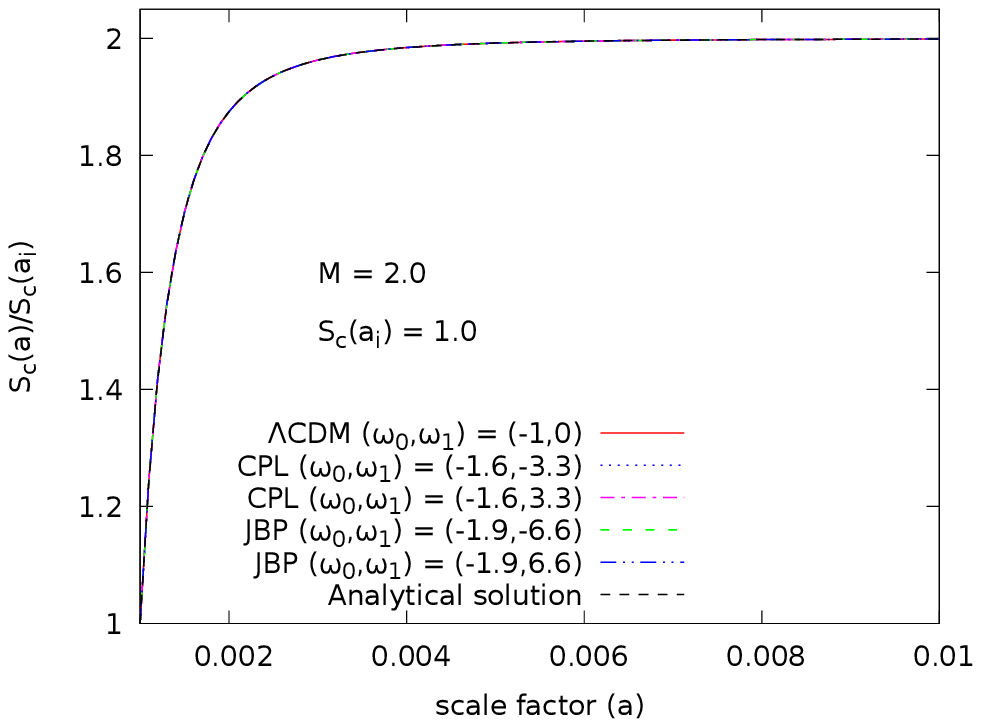}}}
   \hspace{0.5 cm}
   \resizebox{7 cm}{!}{\rotatebox{0}{\includegraphics{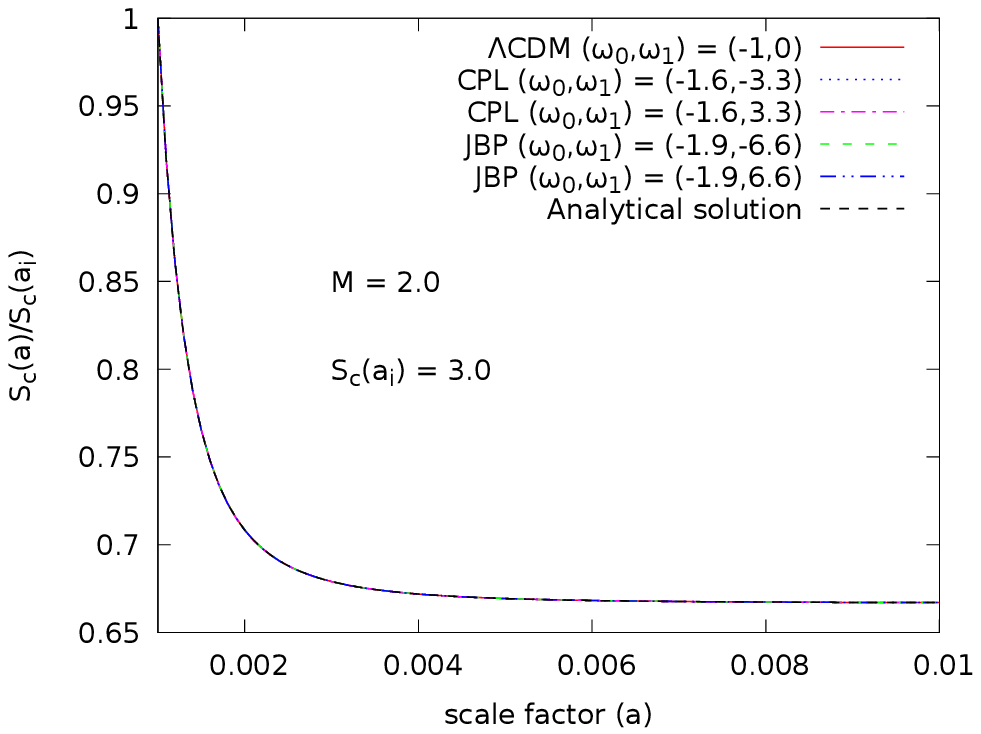}}}\\
   \vspace{-0.1 cm} 
   \resizebox{7 cm}{!}{\rotatebox{0}{\includegraphics{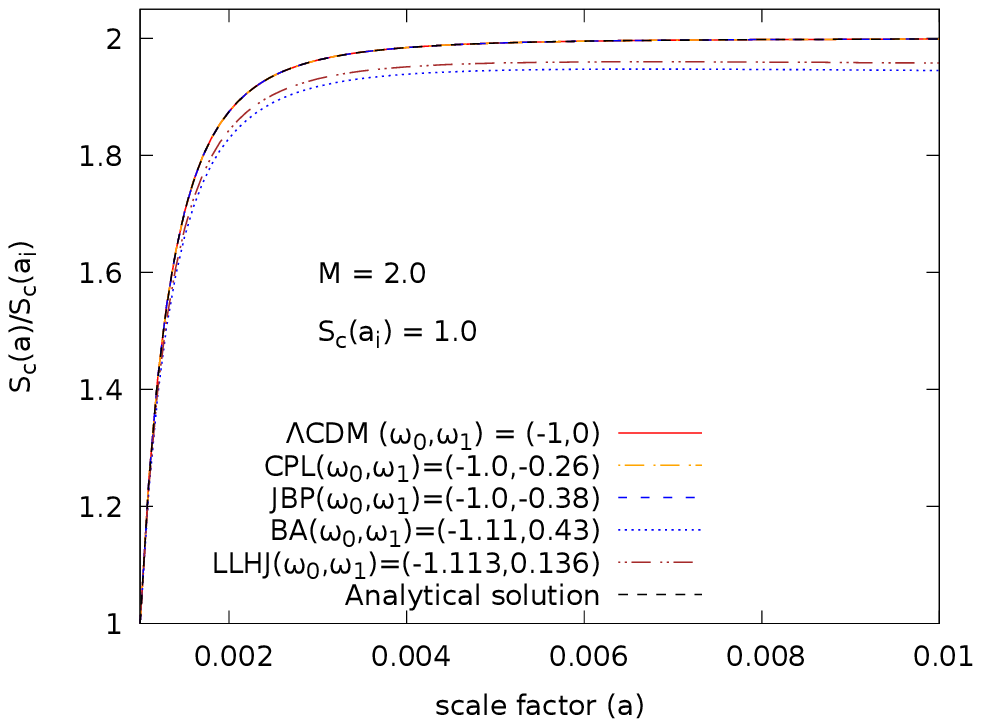}}}
   \hspace{0.5 cm}
   \resizebox{7 cm}{!}{\rotatebox{0}{\includegraphics{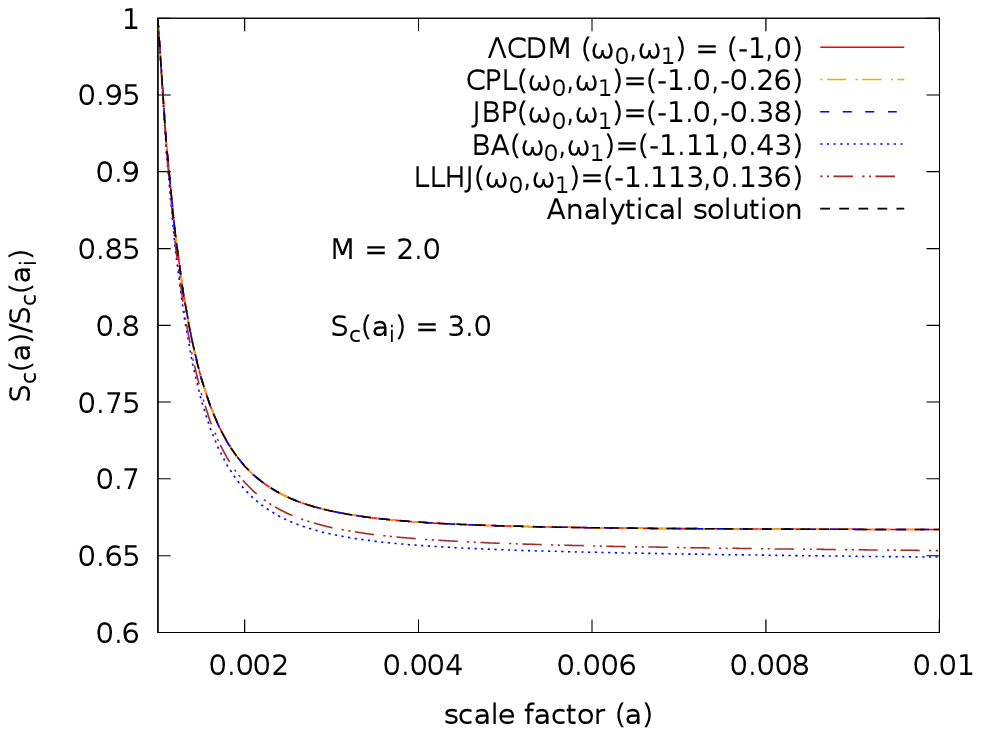}}}\\
 \caption{The top left and top right panels show the initial
     transient in $\frac{S_c(a)}{S_c(a_i)}$ near the initial scale
     factor $a_i$ for CPL and JBP parametrization along with the
     $\Lambda$CDM model for $S_c(a_i)<M$ and $S_c(a_i)>M$
     respectively. The bottom two panels show the same but for BA and
     LLHJ parametrization along with CPL and JBP parametrization with
     different choice of parameters. The approximate analytical
     solution (\autoref{eq:nine}) of \autoref{eq:eight} are also
     plotted in each case for a comparison.} 
   \label{fig:transient}
  \end{figure*}

  We study the evolution of the configuration entropy in different
  cosmological models by solving \autoref{eq:seventeen} and
  \autoref{eq:eight}. The results are shown in different panels of
  \autoref{fig:ent}. We consider three different combinations of the
  initial configuration entropy $S_c(a_i)$ and the mass $M$. In the
  top left panel of \autoref{fig:ent}, we show the configuration
  entropy as a function of the scale factor $a$ in the dynamical dark
  energy models and the $\Lambda$CDM model for $S_c(a_i)=1$ and $M=2$.
  We note that initially there is a sharp increase in the
  configuration entropy in all the models which soon stabilizes to a
  constant value. The configuration entropy then starts decreasing
  with increasing scale factor in both the $\Lambda$CDM model and the
  dynamical dark energy models with $\omega_1<0$. We find that the
  decrease of the configuration entropy with increasing scale factor
  is very similar in the two models with CPL and JBP parameterization
  of dark energy equation of state with $\omega_1<0$. We observe that
  these models with negative $\omega_1$ exhibit a larger decrease in
  the configuration entropy as compared to the $\Lambda$CDM model
  after $a=0.5$. It may be noted in the \autoref{fig:e2d2fwall} that
  these models with negative $\omega_1$ show a significantly higher
  growth rate as compared to the $\Lambda$CDM model which leads to a
  larger dissipation of the configuration entropy in these
  models. These models are the ones in which the dark energy
  domination starts early but it becomes less dominant at late
  times. On the other hand, the dynamical dark energy models with
  $\omega_1>0$ in both the CPL and JBP parametrization do not show a
  decrease in the configuration entropy after the initial increase in
  its value. These are the models in which the dark energy domination
  starts at a later stage suppressing the growth of structures which
  can be clearly seen in the significantly smaller values of the
  growth rate in these models in \autoref{fig:e2d2fwall}.

 The top right panel of \autoref{fig:ent} shows the evolution of the
 configuration entropy in different cosmological models for
 $S_c(a_i)=2$ and $M=2$. The bottom panel shows the same but for the
 combination $S_c(a_i)=3$ and $M=2$. The choice of these parameters
 are somewhat arbitrary here and we consider three representative
 examples each for $S_c(a_i)<M$, $S_c(a_i)=M$ and $S_c(a_i)>M$. It may
 be noted here that in this work, $S_c(a_i)$ and $M$ have the same
 dimension as we do not normalize the density $\rho$ by the total mass
 $M$ in \autoref{eq:one}.

 The top right and the bottom panels of \autoref{fig:ent} show the
 same trends in the evolution of configuration entropy in different
 cosmological models as observed in the top left panel of
 \autoref{fig:ent} but with one remarkable difference. In the top left
 panel, we observe a sharp increase in the configuration entropy near
 the initial value of the scale factor. Contrary to this, the
 configuration entropy shows a sharp decrease near the initial value
 of the scale factor when $S_c(a_i)>M$. This can be clearly seen in
 the bottom panel of \autoref{fig:ent}. No such sharp increase or
 decrease can be seen in the configuration entropy near the initial
 scale factor when we choose $S_c(a_i)=M$. The corresponding results
 are shown in the top right panel of \autoref{fig:ent}. Initially the
 evolution of the configuration entropy is determined by the second
 term in \autoref{eq:eight} as the cosmology dependent third term
 remains negligible at an early stage of evolution. We expect that
 this particular behaviour of the configuration entropy may be
 exploited to constrain its initial value from future observations.

We have also considered a set of models in which the equation of state
parameter $\omega(a)$ does not deviate much from $-1$ in the entire
range of evolution. We show the evolution of the configuration entropy
for different combinations of $S_{c}(a_i)$ and $M$ in these models in
\autoref{fig:entnew}. The \autoref{fig:entnew} shows that the nature
of evolution at the initial stage is primarily decided by the values
of $S_{c}(a_i)$ and $M$, as noticed earlier for the other models in
\autoref{fig:ent}. The CPL and JBP parametrizations with a different
choice of parameters in this case are nearly indistinguishable from
the $\Lambda$CDM model. The differences between the $\Lambda$CDM model
and these models are noticeable after $z>1$ only when $S_c(a_i)=M$ in
the top right panel of \autoref{fig:entnew}. On the other hand, the BA
and LLHJ parametrizations show a quite discernible behaviour compared
to the $\Lambda$CDM model or the CPL and JBP parametrizations
throughout the entire range of evolution. Interestingly, the
configuration entropy in these models decreases with time and then
start increasing again after redshift $\sim 0.5$.

  Finally, we explore the transients observed in the numerical values
  of $\frac{S_c(a)}{S_c(a_i)}$ near the initial scale factor $a_i$
  when $S_c(a_i)<M$ and $S_c(a_i)>M$ (top left and bottom panels of
  \autoref{fig:ent} and \autoref{fig:entnew}). The behaviour of these
  initial transients can be easily explained by the approximate
  analytical solution (\autoref{eq:nine}) of
  \autoref{eq:eight}. Initially, the third term would be negligible as
  compared to the second term in \autoref{eq:eight}. The cosmology
  dependent third term would prevail later only when the growth of
  structures becomes significant. We show the transients near the
  initial scale factor for each of the models in different panels of
  \autoref{fig:transient}. The results for the $\Lambda$CDM model
  along with the CPL and JBP parametrizations are shown in the top two
  panels of \autoref{fig:transient}. The top left and the top right
  panel show the behaviour of the transient for $S_c(a_i)<M$ and
  $S_c(a_i)>M$ respectively. We expect a sudden growth in the
  configuration entropy near the initial scale factor when
  $S_c(a_i)<M$. The second term in the analytical solution
  (\autoref{eq:nine}) then becomes negative and consequently entropy
  rises proportional to $a^{-3}$. On the other hand, $S_c(a_i)>M$
  leads to a sudden decay in the entropy near the initial value of the
  scale factor. We also compare the numerical solutions of
  \autoref{eq:eight} for different models with its approximate
  analytical solution near the initial scale factor. In the top two
  panels of \autoref{fig:transient}, we find that the transients
  observed in the numerical solutions in each model are well described
  by the approximate analytical solution. The bottom two panels of
  \autoref{fig:transient} show the transients in the BA and LLHJ
  parametrization along with CPL and JBP parametrization with
  different choice of parameters. It is interesting to note that both
  the LLHJ and BA parametrizations show noticeable deviation from the
  approximate analytical results starting from the very small value of
  scale factor ($a\sim 0.002$). This indicates that the cosmology
  dependent term in \autoref{eq:eight} becomes relevant even at an
  early stage of evolution in these two models. Earlier, we notice in
  \autoref{fig:entnew} that $\frac{S_c(a)}{S_c(a_i)}$ show a
  qualitatively similar behaviour in the LLHJ and BA parametrizations
  despite their different equation of states and growth rates. However
  it may be noted that it is the product of $f$ and $D^2(a)$ in the
  third term of \autoref{eq:eight} which determines the behaviour of
  entropy in a particular model.

The present analysis deals with the evolution of the configuration
entropy in the linear regime. It would be also interesting to study
its behaviour in the non-linear regime. The anisotropic gravitational
collapse of the overdense regions leads to different types of
non-linear structures such as sheets, filaments and clusters. The
evolution of the density field and the velocity field become quite
diverse in these environments. Consequently, it becomes difficult to
capture the evolution with a simple analytic framework. However, one
can track the evolution of the configuration entropy in the non-linear
regime using N-body simulations.

In the present work, we have considered a set of different models for
dark energy and explore the evolution of the configuration entropy in
these models. The evolution of the configuration entropy is primarily
governed by the growth rate of structures and the expansion rate of
the Universe. The differences in the configuration entropy as shown in
\autoref{fig:ent} and \autoref{fig:entnew} arise due to the different
expansion history and growth rate in different models. Our results
suggest that the models with $\omega_1>0$ and $\omega_1<0$ can be
easily distinguished from their distinct behaviour with respect to
each other. It is interesting to note that the nature of evolution of
the configuration entropy also depends on the adopted
parametrization. So one may also discern the dynamical dark energy
models with different parametrizations by studying the evolution of
the configuration entropy. We hope that an analysis of the
configuration entropy from future observations may enable us to
constrain these models in an efficient manner.

\section {Acknowledgement}
We sincerely thank an anonymous referee for useful comments and
suggestions which helped us to improve the draft. The authors would
like to thank Eric V. Linder, Sudipta Das and Suman Sarkar for useful
discussions. The authors would also like to acknowledge financial
support from the SERB, DST, Government of India through the project
EMR/2015/001037. BP would also like to acknowledge IUCAA, Pune and
CTS, IIT, Kharagpur for providing support through associateship and
visitors programme respectively.

\bsp	
\label{lastpage}

\begin{thebibliography}{99}

\bibitem[Armendariz-Picon et al.(2001)]{armendariz} Armendariz-Picon,
  C., Mukhanov, V., \& Steinhardt, P.~J.\ 2001, \prd, 63, 103510
  
\bibitem[Amendola \& Tsujikawa(2010)]{de2010} Amendola, L. \& Tsujikawa, S.\ 2010
  Dark Energy: Theory and Observation, Cambridge University Press

\bibitem[Barboza \& Alcaniz(2008)]{barboza} Barboza, E.~M., \& Alcaniz, J.~S.\ 2008, Physics Letters B, 666, 415

\bibitem[Bekenstein(1973)]{bekenstein} Bekenstein, J. D.\ 1973,
  \prd, 7, 2333

\bibitem[Buchert(2000)]{buchert2k} Buchert, T.\ 2000, General
  Relativity and Gravitation, 32, 105

\bibitem[Caldwell et al.(1998)]{caldwell} Caldwell, R.~R., Dave, R.,
  \& Steinhardt, P.~J.\ 1998, Physical Review Letters, 80, 1582

\bibitem[Chevallier \& Polarski(2001)]{chevallier} Chevallier, M., 
  \& Polarski, D.\ 2001, International Journal of Modern Physics D, 10, 213

\bibitem[Copeland et al.(2006)]{copeland} Copeland, E.~J., Sami, M.,
  \& Tsujikawa, S.\ 2006, International Journal of Modern Physics D,
  15, 1753

\bibitem[Easson et al.(2011)]{easson} Easson, D.~A., Frampton, P.~H.,
  \& Smoot, G.~F.\ 2011, Physics Letters B, 696, 273

\bibitem[Egan \& Lineweaver(2010)]{egan} Egan, C.~A., \& Lineweaver,
  C.~H.\ 2010, \apj, 710, 1825

\bibitem[Ferreira \& Pav{\'o}n(2016)]{ferreira} Ferreira, P.~C., \&
  Pav{\'o}n, D.\ 2016, European Physical Journal C, 76, 37

\bibitem[Frampton(2009)]{frampton} Frampton, P.~H.\ 2009, \jcap, 10,
  016

\bibitem[Gibbons \& Hawking(1977)]{gibbons} Gibbons, G.~W., \&
  Hawking, S.~W.\ 1977, \prd, 15, 2738

\bibitem[Gong(2005)]{gong} Gong, Y.\ 2005, Classical and Quantum Gravity, 22, 2121 

\bibitem[Hawking(1976)]{hawking} Hawking, S. W.\ 1976, \prd, 13, 191

\bibitem[Hubble (1929)]{hubble} Hubble, E. \ 1929, 
Proceedings of the National Academy of Sciences of the U.S.A., 15, 168
  
\bibitem[Hunt \& Sarkar(2010)]{hunt} Hunt, P. \& Sarkar, S. \ 2010,
  \mnras, 401, 547

\bibitem[Jassal et al.(2005)]{jbp} Jassal, H.~K., Bagla, J.~S., 
  \& Padmanabhan, T.\ 2005, \mnras, 356, L11 

\bibitem[Johri \& Rath(2006)]{johri} Johri, V.~B., \& 
  Rath, P.~K.\ 2006, \prd, 74, 123516 
 
\bibitem[Linder(2003)]{linder1} Linder, E.~V.\ 2003, Physical Review Letters, 90, 091301  
 
\bibitem[Linder(2005)]{linder2} Linder, E.~V.\ 2005, \prd, 72, 043529


\bibitem[Linder \& Jenkins(2003)]{linder4} Linder, E.~V., \& Jenkins, A.\ 2003, \mnras, 346, 573 

\bibitem[Liu et al.(2008)]{liu} Liu, D.-J., Li, X.-Z., Hao, J., \& Jin, X.-H.\ 2008, \mnras, 388, 275

\bibitem[Mamon et al.(2018)]{mamon} Mamon, A.~A., Bhandari, P., \& Chakraborty, S.\ 2018, arXiv:1802.07925 

\bibitem[Milton(2003)]{milton} Milton, K.~A.\ 2003, Gravitation and Cosmology, 9, 66 

\bibitem[Mimoso \& Pav{\'o}n(2013)]{mimoso} Mimoso, J.~P., \&
  Pav{\'o}n, D.\ 2013, \prd, 87, 047302

\bibitem[Padmanabhan(2017)]{paddy} Padmanabhan, T.\ 2017, Comptes Rendus Physique, 18, 275,

\bibitem[Padmanabhan \& Padmanabhan(2017)]{paddyhamsa}
  Padmanabhan, T., \& Padmanabhan, H.\ 2017, Physics Letters B, 773,
  81
\bibitem[Pandey(2017)]{pandey1} Pandey, B.\ 2017, \mnras, 471, L77

\bibitem[Pav{\'o}n \& Radicella(2013)]{pavon1} Pav{\'o}n, D., \&
  Radicella, N.\ 2013, General Relativity and Gravitation, 45, 63

\bibitem[Pav{\'o}n(2014)]{pavon2} Pav{\'o}n, D. \ 2014, International
  Journal of Geometric Methods in Modern Physics, , 11, 1460007

\bibitem[Penrose(2004)]{penrose} Penrose, R.\ 2004, The road to
  reality : a complete guide to the laws of the universe.~ London: Jonathan Cape, 2004,

\bibitem[Perlmutter et al.(1999a)]{perlmutter1} Perlmutter, S.,
  Aldering, G., Goldhaber, G., et al.\ 1999, \apj, 517, 565

\bibitem[Radicella \& Pav{\'o}n(2012)]{radicella} Radicella, N., \&
  Pav{\'o}n, D.\ 2012, General Relativity and Gravitation, 44, 685

\bibitem[Ratra \& Peebles(1988)]{ratra} Ratra, B., \& 
  Peebles, P.~J.~E.\ 1988, \prd, 37, 3406 

\bibitem[Riess et al.(1998)]{riess} Riess, A.~G.,
  Filippenko, A.~V., Challis, P., et al.\ 1998, \aj, 116, 1009

\bibitem[Shannon(1948)]{shannon48} Shannon, C. E. \ 1948, Bell
System Technical Journal, 27, 379-423, 623-656

\bibitem[Tripathi et al.(2017)]{tripathi} Tripathi, A., Sangwan, A., \& Jassal, H.~K.\ 2017, \jcap, 6, 012

\bibitem[Tsujikawa(2010)]{tsujikawa} Tsujikawa, S.\ 2010, Lecture
  Notes in Physics, Berlin Springer Verlag, 800, 99

\bibitem[Tomita(2001)]{tomita01} Tomita, K. \ 2001, \mnras, 326, 287

\bibitem[York et al.(2000)]{york} York, D.~G., et al.\ 2000, \aj,
  120, 1579

\end{thebibliography}
\end{document}